\RequirePackage{ifpdf}
\documentclass[11pt,a4paper]{article}
\usepackage{epsfig,jheppub}

\usepackage{multirow,longtable,enumerate}

\usepackage{physics}
\usepackage{tensor}
\usepackage{amsmath}
\usepackage{amssymb}
\usepackage{appendix}
\usepackage{mathtools}
\usepackage{mathrsfs}
\usepackage[normalem]{ulem}
\usepackage{tikz}
\usepackage{booktabs}
\usepackage{bbm}
\usepackage{verbatim}

\font\cmss=cmss12

\newcommand{\mN}{\mathcal N}
\newcommand{\mA}{\mathcal A}
\newcommand{\mB}{\mathcal B}
\newcommand{\bo}{\bar{1}}
\newcommand{\bt}{\bar{2}}

\newcommand\half{\frac12}
\newcommand\del{\partial}
\newcommand\bi{\begin{itemize}}
\newcommand\ei{\end{itemize}}

\newcommand\tc{{\tilde c}}

\newcommand\tilh{{\tilde h}}
\newcommand\tchi{{\widetilde \chi}}

\newcommand\mS{\mathcal{S}}

\newcommand\bea{\begin{eqnarray}}
\newcommand\eea{\end{eqnarray}}
\newcommand\be{\begin{equation}}
\newcommand\ee{\end{equation}}

\newcommand\btau{{\bar \tau}}
\newcommand\cA{{\cal A}}
\newcommand\cB{{\cal B}}
\newcommand\cT{{\cal T}}

\newcommand\cN{{\cal N}}

\newcommand\cD{{\cal D}}

\newcommand\bchi{{\overline \chi}}

\newcommand\sfrac[2]{{\textstyle\frac{#1}{#2}}}

\newcommand\ZZ{\hbox{Z\kern-.4emZ}}
\newcommand\sZZ{\hbox{\sevenfont Z\kern-.4emZ}}

\newcommand{\cM}{{\cal M}}

\newcommand{\eref}[1]{Eq.\,(\ref{#1})}
\newcommand{\Comment}[1]{{}}

\newcommand{\wtd}{\widetilde}
\newcommand{\mc}{\mathcal}

\newcommand{\bbo}{\mathbbm{1}}
\allowdisplaybreaks

\def\IB{\relax{\rm I\kern-.18em B}}
\def\IC{{\relax\hbox{\kern.3em{\cmss I}$\kern-.4em{\rm C}$}}}
\def\ID{\relax{\rm I\kern-.18em D}}
\def\IE{\relax{\rm I\kern-.18em E}}
\def\IF{\relax{\rm I\kern-.18em F}}
\def\II{\relax{\rm I\kern-.18em I}}

\def\Id{\relax{1\kern-.32em 1}}
\def\IG{\relax\hbox{$\inbar\kern-.3em{\rm G}$}}
\def\IR{\relax{\rm I\kern-.18em R}}

\title{Rational CFT With Three Characters: The Quasi-Character Approach} \author{Sunil Mukhi\footnote{Email: sunil.mukhi@gmail.com}, Rahul Poddar\footnote{Email: rahul.poddar.305@gmail.com} and Palash Singh\footnote{Email:
palash13singh@gmail.com}\\ \it Indian Institute of Science Education and Research,\\ \it Homi Bhabha Rd, Pashan, Pune 411 008, India
} 

\abstract{Quasi-characters  are vector-valued modular functions having an integral, but not necessarily positive, $q$-expansion. Using modular differential equations, a complete classification has been provided in \href{https://arxiv.org/abs/1810.09472}{arXiv:1810.09472} for the case of two characters. These in turn generate all possible admissible characters, of arbitrary Wronskian index, in order two. Here we initiate a study of the three-character case. We conjecture several infinite families of quasi-characters and show in examples that their linear combinations can generate admissible characters with arbitrarily large Wronskian index. The structure is completely different from the order two case, and the novel coset construction of \href{https://arxiv.org/abs/1602.01022}{arXiv:1602.01022} plays a key role in discovering the appropriate families. Using even unimodular lattices, we construct some explicit three-character CFT corresponding to the new admissible characters.}

\preprint{}

\keywords{Conformal field theory, Modular invariance, Conformal bootstrap}

\arxivnumber{}

\begin{document}

\maketitle

\section{Introduction}

The classification of characters for 2d rational conformal field theories (RCFT) is a problem that is both interesting and somewhat tractable, particularly for a small number of characters. For meromorphic (one-character) CFT, the classification of admissible characters is a solved problem: they have to be specific types of functions of the Klein $j$-invariant. But for $p\ge 2$ characters, the problem is non-trivial. Techniques used towards this end include modular linear differential equations (MLDE) \cite{Mathur:1988na,Mathur:1988gt, Naculich:1988xv, Kiritsis:1988kq, Kiritsis:1989kc, Durganandini:1989es, Kaneko:2003, Kaneko:2006, Mason:2007, Mason:2008,Tuite:2008pt,Kaneko:2013uga, Hampapura:2015cea, Gaberdiel:2016zke, Hampapura:2016mmz, Arike:2016ana, Franc:2016, Tener:2016lcn,Mason:2018,Chandra:2018pjq, Franc:2019}, fusion algebras \cite{Christe:1988xy, Mathur:1989pk, Ostrik:2002}, representation theory of SL(2,$\mathbb Z$) \cite{Bantay:2005vk,Bantay:2007zz,Gannon:2013jua}, Hecke operators \cite{Harvey:2018rdc}, and contour-integral representations \cite{Mukhi:1989qk,Mukhi:2019cpu}. 

Crucially these classification procedures do not assume any particular chiral algebra, and do not privilege the Virasoro algebra. Rather, they start by specifying the order -- the number of characters\footnote{This is the dimension of the vector-valued modular function, as well as the order of the associated differential equation. We will consistently use ``order'' to denote the number of characters.}, as well as a quantity called the Wronskian index. The chiral algebra as well as the precise number of primary fields under it (which can be larger than the number of characters) are outputs of the classification method. The modular $\mS$-matrix and fusion rules, as well as correlation functions, are also outputs. By this process one can in many cases reconstruct the entire RCFT \cite{Mathur:1988gt}. 

Candidate characters for CFT must be holomorphic in moduli space except at infinity, and have three more essential properties: (i) they transform under modular transformations as vector-valued modular functions, so that the partition function is modular invariant, (ii) the ground state in the identity character is non-degenerate, so that the ground state of the full CFT is non-degenerate, as should be the case in any legitimate quantum field theory, (iii) each character has non-negative integer degeneracies of states after suitably normalising the ground state, except for the identity which is normalised to unity. Objects with these three properties are called ``admissible'' characters. While some admissible characters correspond to more than one CFT, others do not correspond to any CFT\footnote{This is the significance of the seminal work \cite{Schellekens:1992db} for the meromorphic case. The analogous question in order $p=2$ was studied in \cite{Chandra:2018ezv}.}. Clearly this approach to classifying RCFT is quite different from the more familiar one where a chiral algebra is chosen and then a minimal series constructed by decoupling null vectors \cite{Belavin:1984vu,Knizhnik:1984nr}.

For order 2, a complete classification of admissible characters was provided for the first time in \cite{Chandra:2018pjq}. This classification relies on an auxiliary construction called ``quasi-characters''. These share most of the properties of admissible characters, but are allowed to have some negative coefficients in the $q$-expansion. While quasi-characters do not directly correspond to any CFT, their linear combinations with semi-definite coefficients were shown to generate all admissible characters in order 2. In \cite{Chandra:2018ezv}, many of these linear combinations were then shown to correspond to actual CFT.

In the present work, we conjecture infinitely many families of quasi-characters in order 3. The results have very different features from those in order 2. In particular, the novel coset construction of \cite{Gaberdiel:2016zke} plays a crucial role in identifying these families. By studying their $q$-series to large orders we provide strong support for our conjectures. Moreover, pairs of quasi-characters are found to satisfy the bilinear relation, a hallmark of the novel coset construction, which we again verify to high orders. This fact provides necessary information about degeneracies, as we will see. The resulting picture contrasts sharply with the order-2 case where there are just four families of quasi-characters. The completeness of our list in order 3 is yet to be established and we will leave this for future work.

The outline of this paper is as follows. In Section \ref{review} we briefly review the MLDE approach and then the construction in \cite{Chandra:2018pjq} of quasi-characters of order 2, as well as the method of taking linear combinations to generate admissible characters. In Section \ref{three.quasi} we present our results on quasi-characters of order 3. Then in Section \ref{three.admiss} we show, through examples, that linear combinations (with suitably bounded coefficients) can generate admissible characters, and present examples of genuine CFT that correspond to some of these. Thereafter we summarise our work and list interesting future directions. In the Appendix we list all the cases and the order in $q$ to which we have explicitly confirmed our various conjectures.

\section{Review of background material}

\label{review}

\subsection{Modular linear differential equations (MLDE)}

Admissible characters should transform as vector-valued modular functions of some dimension $p\ge 1$ under the SL(2,$\mathbb Z$) transformations of the modular parameter $\tau$:
\be
\begin{split}
\chi_i(\tau+1)&=e^{2\pi i\alpha_i}\chi_i(\tau)\\
\chi_i\left(-\frac{1}{\tau}\right) &=\sum_j \mS_{ij}\chi_j(\tau)
\end{split}
\ee
where $i=0,1,\cdots,p-1$. 
They should take the form:
\be
\chi_i(\tau) = q^{\alpha_i}\left(a_0^{(i)} + a_1^{(i)} q + a_2^{(i)} q^2 + \cdots\right)\\
\label{seriesexp}
\ee
where $q=e^{2\pi i\tau}$. The exponents $\alpha_i$ are related to the central charge and holomorphic conformal dimensions as:
\be
\alpha_0=-\frac{c}{24},\quad \alpha_i=-\frac{c}{24}+h_i,~i=1,2,\cdots p-1 
\label{alphach}
\ee
The coefficients $a_n^{(i)}$ are non-negative integers, with $a_0^{(0)}=1$. They can then, in principle, be interpreted as degeneracies of secondary states under some chiral algebra. The corresponding partition function:
\be
Z(\tau,\btau)=\sum_{i=0}^{p-1} M_i\, \bchi_i(\btau)\chi_i(\tau)
\ee
will then be modular invariant for suitably chosen integers $M_i$, corresponding to multiplicities of primaries with the same character, as long as suitable conditions are satisfied by the modular transformation matrix. 

It is easily shown that such vector-valued modular functions arise as the independent solutions of a holomorphic and modular-invariant linear differential equation (MLDE) of order $p$. Accordingly, one may search for them by classifying MLDE and their admissible solutions. A general MLDE can be written in monic form as:
\be
\left(\cD^p + \sum_{k=0}^{p-1} \phi_{2(p-k)}(\tau)\cD^k\right)\chi=0
\label{MLDEgeneral}
\ee
where each $\phi_{2j}$ is a modular function of weight $2j$ and $\cD$ is the covariant derivative on moduli space $\frac{1}{2\pi i}\del_\tau - \frac{r}{12} E_2$, acting on weight $r$ modular forms, where $E_2(\tau)$ is the second Eisenstein series. The $p$ independent solutions of the above equation are manifestly vector-valued modular functions. The goal is then to find out for which values of the parameters these solutions are admissible as defined above.

For a fixed order $p$, MLDE's are classified by an additional piece of data, namely the Wronskian index $\ell$ of the characters. This is defined as the integer such that $\frac{\ell}{6}$ is the number of zeroes of the Wronskian determinant of the independent solutions:
\be
W_p\equiv 
\begin{pmatrix}
\chi_0 & \cdots &  \chi_{p-1}\\
D \chi_0 &\cdots &D \chi_{p-1}\\
\vdots & &\vdots \\
 D^{p-1}\chi_0 &\vdots &D^{p-1}\chi_{p-1}\\
\end{pmatrix}
\ee
Equivalently, $\frac{\ell}{6}$ is the maximum number of poles that the $\phi_{2j}$ are allowed to have. The number is fractional due to the orbifold nature of moduli space. It is known that this permits all non-negative integers $\ell$ other than 1. For any fixed value of $\ell$, the $\phi_{2j}$ are determined in terms of a finite number of parameters \cite{Mathur:1988na, Mathur:1988gt}. Almost all of the literature in this area has been focused on the $\ell=0$ case, the only case where the MLDE in monic form is holomorphic. However there is no reason to consider this a ``typical'' case. One of the fascinating developments of the last few years has been the generation of interesting results beyond $\ell=0$ \cite{Hampapura:2015cea,Gaberdiel:2016zke,Harvey:2018rdc,Chandra:2018pjq,Chandra:2018ezv,Harvey:2019qzs}.

By substituting the series expansion \eref{seriesexp} into the MLDE, one finds first of all an indicial equation that relates the parameters of the equation with the critical exponents $\alpha_i$. Higher orders in the series solution then determine the ratios $\frac{a_n^{(i)}}{a_0^{(i)}}$  recursively. In general these are rational functions of the input parameters with growing denominators. However for some values of the parameters the coefficients are rational with denominators that stabilise for large orders of $q$ (except for the identity where they must be integer from the outset), as well as non-negative. By absorbing the stable denominators into the normalisation of the non-identity characters, we find admissible characters which are candidates for an RCFT. We re-label the normalisation $a_0^{(i)}$ as $D_i$ for future discussions. Being linear and homogeneous, the MLDE cannot determine $D_i$. One has to use other methods to do so. This is an essential part of the task of finding admissible characters.

For MLDE of order 2 it can be shown that $\ell$ is an even integer \cite{Naculich:1988xv}. For the values $\ell=0,2,4$  the MLDE approach is fruitful and one finds small sets of solutions in each case. However, no solutions for $\ell\ge 6$ have been generated in the MLDE approach, since the number of parameters in the equation grows too fast to be tractable. Fortunately there is an alternate approach \cite{Chandra:2018pjq} that allows us to investigate these cases {\em without} the use of $\ell\ge 6$ MLDE. We now turn to a description of this approach.

\subsection{Quasi-characters in order 2}

It turns out that for $\ell=0,2,4$ the same second-order MLDE's that give rise to finitely many admissible characters actually have infinitely many solutions if one relaxes the admissibility criterion to allow {\em negative} integral coefficients in the $q$-series. Such solutions are called quasi-characters. Inspired by the mathematical works of \cite{Kaneko:1998,Kaneko:2003,Kaneko:2006}, it was shown in  \cite{Chandra:2018pjq} that they fall into various infinite families, one for each fusion class. The admissible characters occur within these families as special cases with all non-negative coefficients, while the remaining members have at least one negative coefficient in their $q$-series expansion. By taking appropriate linear combinations of quasi-characters within a particular fusion class, one can sometimes cancel the negative signs and find admissible characters. 

Starting with quasi-characters of Wronskian index $\ell=0,2,4$, such linear combinations turn out to have $\ell=6m,6m+2,6m+4$ respectively, where $m$ is an arbitrary positive integer. This method generates infinite families of admissible characters spanning all allowed values of $\ell$. Finally it has been proved that this classification is complete: any admissible character with $\ell=6m$ can be expressed as a linear combination of $\ell=0$ quasi-characters, with similar results holding in the other cases. 

We now review the quasi-character construction of \cite{Chandra:2018pjq}. For our present purposes it will be sufficient to review the $\ell=0$ case, though in this work they have been constructed for $\ell=0,2,4$.

Let us first establish some background and notation. It is known \cite{Christe:1988xy} that all two-character RCFT fall into a small, finite set of fusion classes. Indeed, when there are exactly two primary fields (including the identity) one can have two possible fusion classes, labelled $\cA_1^{(1)}$ and $\cA_1^{(2)}$ in the above reference (the subscript is the number of primaries other than the identity). One can also have three primary fields of which two are a complex-conjugate pair having the same character. For this case there is a unique fusion class, denoted $\cB_2^{(1)}$. Finally, there is a case with four primary fields of which three are related by triality, this class is denoted $\cA_3^{(1)}$. In \cite{Chandra:2018pjq} these fusion classes: $\cA_1^{(1)},\cA_1^{(2)},\cB_2^{(1)},\cA_3^{(1)}$ were re-labelled $A_1$, Lee-Yang, $A_2,D_4$ respectively. We will use this notation below.

For the special case of Wronskian index $\ell=0$, the MLDE is \cite{Mathur:1988na}:
\begin{equation}
  \left( {\cal D}^2  +\mu E_4\right)\chi(\tau) = 0 \label{MLDE2}
\end{equation}
The key observation is that infinite families of quasi-characters of order two exist for the following sets of central charges, labelled by their fusion class:
\be
\begin{split}
\hbox{Lee-Yang:}\qquad c &=\sfrac25(6k+1),\quad k\ne 4 \hbox{ mod }5\\
A_1\!:\qquad c &=6k+1\\
A_2\!:\qquad  c &=4k+2,\quad k\ne 2 \hbox{ mod } 3\\
D_4\!:\qquad c&=12k+4
\end{split}
\label{c.quasi}
\ee
where $k\in\mathbb Z$. 
We can determine the corresponding value of $h$ in each case from the Riemann-Roch theorem. For a general RCFT this tells us that:
\be
\sum_{i=0}^{p-1}\alpha_i=\sum_{i=0}^{p-1}\left(-\frac{c}{24}+h_i\right)=\frac{p(p-1)}{12}-\frac{\ell}{6}
\label{Riemann.Roch}
\ee
For the special case of two characters, this reduces to:
\be
-\frac{c}{12}+h=\frac{1-\ell}{6}
\label{Riemann.Roch.2}
\ee
Since we are considering quasi-characters with $\ell=0$, we find that
$h=\frac{c}{12}+\frac{1}{6}$, which determines $h$ for all the cases above. 

While one can take $k$ in \eref{c.quasi} to be any positive or negative integer, the sub-series for negative $k$ just repeats the one for $k\ge 0$ under exchange of the identity and non-identity character: $c\to c-24h,h\to -h$. To be precise, under this exchange we find $k\to -k-2$ for the Lee-Yang and $A_2$ cases, and $k\to -k-1$ for the $A_1$ and $D_4$ cases. Also in the Lee-Yang and $A_2$ cases, $k=-1$ is a forbidden value as can be seen from the above equation. Thus in all cases it suffices to take $k$ to be a non-negative integer.

It should be mentioned that the existence of these quasi-characters can be observed at an ``experimental'' level using MLDE (i.e. in many examples whose degeneracies are computed to high orders in $q$), but it has also been proved using recursion relations and other mathematical results established in \cite{Kaneko:1998,Kaneko:2003,Kaneko:2006}.

Let us highlight some key features of the objects in the above list. For this purpose we start with the first class above (Lee-Yang) and note that when $k$ shifts by 10, the central charge jumps by 24. Accordingly we divide the given set into subsets as:
\be
\begin{split}
\hbox{Lee-Yang (I):}~~k&=10j,10j+1,10j+2,10j+3\\
\hbox{Lee-Yang (II):}~~k&=10j+5,10j+6,10j+7,10j+8
\end{split}
\label{c.LY}
\ee 
From Eqs.(\ref{c.quasi}),(\ref{c.LY}) we  notice that when $j$ jumps by 1, the central charge jumps by 24. Thus, Lee-Yang quasi-characters can be classified into families where $c$ jumps by 24 within successive members of each family.

In \cite{Chandra:2018pjq}, the first subset of quasi-characters was denoted type I. These have negative coefficients in their first few terms, but beyond some order all coefficients are positive. In the second subset one instead finds type II quasi-characters. For these, if the leading term is chosen positive then beyond some order all the coefficients are {\em negative}. These properties hold for the identity character, while the character corresponding to the nontrivial primary has all non-negative coefficients\footnote{\label{unpres} As noted already in \cite{Mathur:1988na}, there is a subtlety that we sometimes exchange the role of identity and non-identity characters in order to achieve non-negative fusion rules. These choices were referred to in \cite{Chandra:2018pjq} as unitary and non-unitary presentations. In the present discussion we always consider the unitary presentation for quasi-characters, because then our statements are uniformly valid in all cases. After constructing admissible characters one will still be free to switch to the non-unitary presentation if desired.}.

An additional property of type I quasi-characters is that the number of negative coefficients is found to be equal to $j$. Thus for $j=0$, we actually find admissible characters, i.e. no negative coefficients. For $j=1$ there is a single negative coefficient that occurs right near the beginning of the $q$-series and thereafter all coefficients are found to be positive, and so on. We will shortly discuss the utility of type I quasi-characters. So far no comparable utility has been found for those of type II.
 
Let us repeat the above  classification into subsets for the remaining lines of \eref{c.quasi}. We find:
\be
\begin{split}
A_1 \hbox{ (I):}~~k&=4j,4j+1\\
A_1 \hbox{ (II):}~~k&=4j+2,4j+3\\[2mm]
A_2 \hbox{ (I):}~~k&=6j,6j+1\\
A_2 \hbox{ (II):}~~k&=6j+3,6j+4\\[2mm]
D_4 \hbox{ (I):}~~k&=2j\\
D_4 \hbox{ (II):}~~k&=2j+1
\end{split}
\label{c.rest}
\ee 
The properties are very analogous to those we discussed for the Lee-Yang case above. Again the quasi-characters fall into sub-families such that each time $j$ jumps by 1, their central charge jumps by 24. Also the cases labelled (I) all correspond to type I quasi-characters and the number of negative coefficients turns out to be $j$. 

The type I cases at $j=0$ in Eqs.(\ref{c.LY}),(\ref{c.rest}) correspond to admissible charaacters, indeed these are precisely the MMS series of rational CFT's \cite{Mathur:1988na}. It has long been known that these are the only admissible characters in order 2 with Wronskian index $\ell=0$. Note however that unlike the admissible characters, quasi-characters can have non-trivial ground-state degeneracies even for the identity character. These can be computed using the method discussed in \cite{Mukhi:2019cpu}.

Now we can exhibit the utility of type I quasi-characters. From any of the examples above, pick all pairs of the same type with $j=0$ and $j=1$. In the Lee-Yang family there are four such pairs with $(c,c')=(\frac25,\frac{122}{5}), (\frac{14}{5},\frac{134}{5}), (\frac{26}{5},\frac{146}{5}),(\frac{38}{5},\frac{158}{5})$. In the $A_1$ family we have pairs with $(c,c')=(1,25),(7,31)$, for $A_2$ we have $(c,c')=(2,26),(6,30)$ and for $D_4$ we have the single pair $(c,c')=(4,28)$. Notice that for each of the nine pairs listed, the first member is an admissible character from the MMS series. Also, in each case $c'=c+24$. It can be checked that the second member of each pair is a type I quasi-character with a single negative coefficient in the second term of the identity character. In other words, this character has the form:
\be
\chi'_0=q^{-\frac{c'}{24}}(1-|a_1|\,q+a_2 q^2+\cdots)
\ee
where all the remaining terms are non-negative. Note now that the term with the negative coefficient corresponds to  the power $q^{-\frac{c'}{24}+1}=q^{-\frac{c}{24}}$. This is precisely the same power as the leading term of the first member of the pair. Thus by adding the identity characters we get:
\be
\tchi_0=\chi'_0+N\chi_0=q^{-\frac{c'}{24}}(1+(N-|a_1|)q+\cdots)
\label{examplesum}
\ee
For $N\ge |a_1|$, the result will be an admissible character. The reason is that all successive terms of the second member $\chi'_0$, as well as all terms of the first member $\chi_0$, are non-negative. Of course we must also add the non-identity characters in the same way:
\be
\tchi_1=\chi'_1+N\chi_1
\label{nonid}
\ee
and this will again be admissible. Moreover -- and this is crucial -- because the central charges for $\chi_i$ and $\chi'_i$ differ by 24, their modular transformation matrices $\cT,\mS$ are the same. Thus the sum taken in this way is a vector-valued modular function of the desired type. 

It remains to work out the central charge and conformal dimension, which we denote $(\tc,\tilh)$, for the sum. Note that given $(c,c')$ we know $(h,h')$ by the Riemann-Roch theorem, since each quasi-character has $\ell=0$. Indeed, we have $h=\frac{c}{12}+\frac{1}{6}$ and $h'=\frac{c'}{12}+\frac{1}{6}=\frac{c}{12}+ \frac{1}{6}+2=h+2$. Now from \eref{examplesum} we see that the sum of the identity characters gives us $\tc=c'$ for the resulting theory. 
The conformal dimension is more subtle. The leading powers of the non-identity characters added in \eref{nonid} are: 
\be
\left(q^{-\frac{c'}{24}+h'}, q^{-\frac{c}{24}+h}\right)=
\left(q^{-\frac{c'}{24}+h'}, q^{-\frac{c'}{24}+h+1}\right)=
\left(q^{-\frac{c'}{24}+h'}, q^{-\frac{c'}{24}+h'-1}\right)
\ee
Of the two, the second power is more singular and tells us that the resulting theory has $\tilh=h'-1$. 

Thus, the admissible characters obtained by the addition above have $(\tc,\tilh)=(c',h'-1)$. From the Riemann-Roch theorem, the Wronskian index of the resulting theory is:
\be
\begin{split}
\ell&=-6\left(-\frac{\tc}{12}+\tilh-\frac{1}{6}\right)\\
&=-6\left(-\frac{c'}{12}+h'-1-\frac{1}{6}\right)\\
&=6
\end{split}
\ee 
So, by adding quasi-characters with $\ell=0$, we have landed on admissible characters with $\ell=6$. This works in all the cases listed above, as long as we restrict ourselves to $j=0,1$ as we did so far. We can generalise this by considering triples with $j=0,1,2$ and adding the three sets of characters. It is straightforward to generalise the above arguments to show that in this case one ends up with $\ell=12$. In general, appropriate sums of $m$ quasi-characters gives us admissible characters with $\ell=6(m-1)$. To complete the classification of admissible characters in order 2, one needs to discuss quasi-characters with $\ell=2,4$ and then take linear combinations. This has been done in \cite{Chandra:2018pjq} but we will not need the results here.  We now move on to the case of order 3, where the results are similar in some ways but strikingly different in others.

\section{Quasi-characters in order 3}

\label{three.quasi}

The classification of admissible characters in order 3, initiated in \cite{Mathur:1988gt}, has yielded very different results from the order 2 case. For example there are infinitely many admissible characters already at $\ell=0$. Additional progress with this classification, leading to the discovery of several new theories, was made in \cite{Tuite:2008pt,Gaberdiel:2016zke,Franc:2019}. However, in order 3 very little has been known about CFT with nonzero Wronskian index $\ell>0$. Given the success of the quasi-character method in order 2, one may ask if it can be generalised to order 3 and used to generate such CFT. For this one needs to investigate whether quasi-characters exist in order 3 and if so, whether their linear combinations generate admissible characters with $\ell>0$.

Below we answer these two questions in the affirmative. We construct doubly infinite families of quasi-characters with $(p=3,\ell=0)$ and show that all known 3-character CFT's with $\ell=0$ occur within these families. Next we take linear combinations of these and demonstrate that, when coefficients are chosen appropriately, the result is an admissible character with $(p,\ell=6m)$ for some positive integer $m$. Then we show the existence of a few actual CFT's corresponding to such $\ell>0$ characters in order 3, leaving a detailed investigation of this question to future work.

In this section we conjecture families of quasi-characters in order 3, analogous to those in Section \ref{review} where the case of order 2 was summarised. We restrict ourselves to the case $\ell=0$, for which any admissible character or quasi-character satisfies:
\be
\sum_{i=0}^2\alpha_i=-\frac{c}{8}+h_1+h_2=\half 
\label{Riemann.Roch.3}
\ee
This is the special case of \eref{Riemann.Roch} for $(p,\ell)=(3,0)$. 

Our conjectures have been verified to high orders in $q$ using MLDE. Details of which cases have been verified, and to what order, can be found in Appendix \ref{verifying}.
 
\subsection{Fusion classes}

Before presenting our results on quasi-characters, we need to review and extend some results on fusion classes. As we saw in order 2 there can, in general, be more primaries than characters. This happens, for example, whenever a primary is complex -- in this case its complex conjugate has the same character. Thus we must be careful to account for multiplicities. For the case of interest to us, namely order 3, it turns out that there are cases with a total of three, four, five or nine primary fields. With three or four primaries there are nine possible fusion classes, classified in \cite{Christe:1988xy}. Two of our cases do not fall within the above classification. For one of them, two of the three characters correspond to complex primaries and hence the total number of primaries (counting each one and its complex conjugate as distinct) is five. We have labelled the fusion rules for this case as $\mB_4^{(1)}$.
Another of our cases is the tensor product of a two-character theory with itself. In this process one obtains 9 primaries. 

The fusion classes are listed here, together with their non-vanishing fusion rule coefficients: 
\be
\begin{split}
  \mA_2^{(1)}: &\quad \mN_{011} = \mN_{022} = \mN_{122} \\
  \mA_2^{(2)}: &\quad \mN_{011} = \mN_{022} = \mN_{122} = \mN_{222} \\
  \mA_2^{(3)}: &\quad \mN_{011} = \mN_{022} = \mN_{112} = \mN_{122} = \mN_{222} \\
  \mA_3^{(1)}: &\quad \mN_{011} = \mN_{022} = \mN_{033} = \mN_{123} \\
  \mA_3^{(4)}: &\quad \mN_{011} = \mN_{022} = \mN_{033} = \mN_{111} = \mN_{123} = \mN_{133} = \mN_{222} = \mN_{233} = \mN_{333} \\
  \mB_3^{(1)}: &\quad \mN_{01\bo} = \mN_{022} = \mN_{111} = \mN_{112} = \mN_{11\bo} = \mN_{1\bo\bo} = \mN_{1\bo2} = \mN_{\bo\bo2} = \mN_{\bo\bo\bo} \\
  \mB_3^{(2)}: &\quad \mN_{01\bo} = \mN_{022} = \mN_{111} = \mN_{122} = \mN_{\bo\bo\bo} = \mN_{\bo22} = \mN_{222} \\
  \mB_3^{(3)}: &\quad \mN_{01\bo} = \mN_{022} = \mN_{111} = \mN_{122} = \mN_{\bo22} = \mN_{\bo\bo\bo} \\
  \mB_3^{(4)}: &\quad \mN_{01\bo} = \mN_{022} = \mN_{112} = \mN_{\bo\bo2} \\
  \mB_4^{(1)}: &\quad \mc N_{01\bo} = \mc N_{02\bt} = \mc N_{11\bt} = \mc N_{\bo\bo2} = \mc N_{122} = \mc N_{\bo\bt\bt}  
\end{split}
\ee
For simplicity we have not listed the 9-primary case, which is discussed in more detail below.

By construction, characters are eigenvectors of the modular $\cT$-transformation $\tau\to\tau+1$ and transform under $\tau\to -\frac{1}{\tau}$ by a non-trivial modular $\mS$-matrix. Their fusion rules can be found using the Verlinde formula \cite{Verlinde:1988sn}:
\begin{equation}
  \mc N_{ijk} = \sum_m \frac{\mc S_{im} \mc S_{jm} \mc S_{km}^*}{\mc S_{0m}} 
\end{equation}

Let us explain the labelling of the fusion classes in \cite{Christe:1988xy}.
The operator product algebra of a CFT having self-conjugate fields $(\phi \times \phi = \bbo)$ with $n$ non-trivial primary fields is denoted $\mc A_n^{(m)}$ (thus the total number of primaries including the identity is $n+1$).
If there are mutually conjugate fields $(\phi \times \bar \phi = \bbo)$ with $n$ non-trivial primaries, then they are denoted $\mc B_n^{(m)}$.
The superscript $(m)$ is just a label to distinguish different fusion classes with the same number of primaries.
Thus, for three character CFT's with unit multiplicity, the relevant fusion classes are $\mc A_2^{(m)}$.
If a CFT has one conjugate pair, then a three character CFT will have four primaries, and be classified into the fusion class $\mc B_3^{(m)}$.
In particular,  the $D_{r,1}$ WZW model for $r\neq 2 \text{ mod } 4$ is in the class $\mc B_3^{(4)}$.
For example the $D_{6,1}$ or SO(12)$_1$ WZW Model has the conjugate pair $\mathbf{32}$ and $\overline{\mathbf{32}}$.
As mentioned above, we have also considered the case SU(5)$_1$ where there are two conjugate pairs, i.e.~five primaries, whose fusion class we have called $\mc B_4^{(1)}$, a case not studied in \cite{Christe:1988xy}.
In this case the conjugate pairs are the $\mathbf 5$, $\overline{\mathbf{5}}$, and $\mathbf{10}$, $\overline{\mathbf{10}}$ respectively. 

The case $A_{2,1} \otimes A_{2,1}$ has 9 primaries since the $A_{2,1}$ theory has fields $\bbo, \mathbf{3},\overline{\mathbf{3}}$, where the $\mathbf{3},\overline{\mathbf{3}}$ are conjugate pairs. 
Thus, the tensor product theory has the 9 primaries:
\begin{equation}
  (\bbo, \bbo), (\bbo, \mathbf{3}), (\bbo, \overline{\mathbf{3}}), (\mathbf{3},\bbo),(\mathbf{3},\mathbf{3}), (\mathbf{3},\overline{\mathbf{3}}), (\overline{\mathbf{3}},\bbo),(\overline{\mathbf{3}},\mathbf{3}), (\overline{\mathbf{3}},\overline{\mathbf{3}})
\end{equation}
The fusion rules for this theory can be computed using the rule $(\phi_1,\phi_2)\times(\phi'_1,\phi'_2)=(\phi_1\times\phi'_1, \phi_2\times\phi'_2)$ together with the fusion rules for
$A_{2,1}$, which are $\mathbf{3}\times\overline{\mathbf{3}}=\bbo$, $\mathbf{3}\times \mathbf{3}=\overline{\mathbf{3}}$. 
We label this fusion class $\mB_8^{(1)}$.

As noted above, fusion rules can be computed if one knows the modular $\mS$-matrix. A general procedure to compute this has been provided by us in \cite{Mukhi:2019cpu} for large classes of characters having $\ell=0$. This class includes, in particular, all $\ell=0$ CFT's of order 3. 
However there is a subtlety in using $\mS$ to compute the fusion rules: where the multiplicity is greater than one, we cannot directly use the Verlinde formula. Instead  we must use the technique in \cite{Mathur:1988gt} to ``expand'' the $\mc S$-matrix to a larger matrix whose dimension is equal to the total number of primaries. This expanded matrix diagonalises the fusion rules of the primaries and can be used in the Verlinde formula.

As an example, consider the case of a 2 character theory where the nontrivial primary has a multiplicity of two.
We repeat the character with multiplicity in the character vector to create a new character vector
\begin{equation}
  \begin{pmatrix}
    \chi_0 \\
    \chi_1
  \end{pmatrix}
  \to \begin{pmatrix}
    \chi_0 \\
    \hat\chi_1 \\
    \hat\chi_2
  \end{pmatrix}
\end{equation}
such that
\begin{equation}
  \begin{pmatrix}
    \chi_0(-\frac{1}{\tau}) \\
    \hat\chi_1(-\frac{1}{\tau}) \\
    \hat\chi_2(-\frac{1}{\tau})
  \end{pmatrix}  = \hat{\mc S}
  \begin{pmatrix}
    \chi_0 (\tau) \\
    \hat\chi_1 (\tau) \\
    \hat\chi_2 (\tau)
  \end{pmatrix}
\end{equation}
The expanded $\hat{\mc S}$-matrix will now have free parameters in its entries, which will be constrained by $\hat\chi_1=\hat\chi_2=\chi_1$ and the symmetry of the expanded $\hat{\mc S}$ matrix.
Plugging in the $\hat{\mc S}$ matrix into Verlinde's formula and demanding the fusion rules to be integral fixes the free parameters of the expanded $\hat{\mc S}$ matrix.
One can do the same thing for characters of order greater than two and non-trivial multiplicity, as long as the multiplicity is small.

\subsection{Quasi-characters}

We would now like to use the $\ell=0$ MLDE to find out whether there exist families of quasi-characters in order 3. To our knowledge this issue has not been addressed in the mathematics literature, unlike the case of order 2 where results of \cite{Kaneko:1998,Kaneko:2003,Kaneko:2006} were used in \cite{Chandra:2018pjq} to provide a classification. 

We therefore use a different approach, based on the novel coset construction of \cite{Gaberdiel:2016zke}. In this work, a series of 15 coset pairs of three-character theories were presented whose central charges add up to 24. The first theory of each pair is a 3-character WZW model, while the second theory is a coset of one of the meromorphic theories classified by Schellekens in \cite{Schellekens:1992db} by the first one. In order 3, the coset of a $c=24$ theory by a three-character theory with $\ell=0$ leads to another CFT which also has $\ell=0$. This is in contrast to the situation in order 2, where the coset by an $\ell=0$ theory leads to a coset with $\ell=2$ (the general formula for the Wronskian index of a coset theory can be found in \cite{Gaberdiel:2016zke}). Now the novel coset relation implies that the modular transformations of a theory and its coset dual are the same. Thus it is reasonable to guess that (in order 3) a theory and its coset dual lie in a common family of quasi-characters, and correspond to the only admissible characters in that family. This guess turns out to be correct, as we have verified in a large number of examples, and and to quite high orders in the $q$-expansion. It immediately generates infinite families of quasi-characters, shown in Table \ref{qcseries}. 
 
\begin{table}[h!]
  \centering
  \begin{tabular}{ccccccc}
    \toprule
    Name & $c$ & $h_1$ & $h_2$ & Fusion Class & Remark \\
    \midrule
    $\mathcal{M}_{2,7}$ & $\frac{304}{7}k - \frac{68}{7}$ & $\frac{20}{7}k - \frac37$ & $\frac97 k - \frac27$ & $\mathcal A_2^{(3)}$ & $k \neq 4$ mod 7 \\[1mm]
    $\mathcal{M}_{3,4}$ & $23k + \frac12$ & $\frac{15}{8}k + \frac{1}{16}$ & $k + \frac12$ & $\mathcal A_2^{(1)}$ & \\[1mm]
    $\mathcal{M}_{2,5} \otimes \mathcal{M}_{2,5}$ & $\frac{208}{5}k - \frac{44}{5}$ & $\frac{14}{5}k - \frac25$ & $\frac{12}{5} k - \frac15$ & $\mA_3^{(4)}$ & $k \neq 3$ mod 5 \\[1mm]
    $G_{2,1} \otimes G_{2,1}$ & $\frac{64}{5}k + \frac{28}{5}$ & $\frac65k + \frac25$ & $\frac25 k + \frac45$ & $\mA_3^{(4)}$ & $k \neq 3$ mod 5 \\[1mm]
    $F_{4,1} \otimes F_{4,1}$ & $\frac{16}{5}k + \frac{52}{5}$ & $\frac45k + \frac35$ & $-\frac25 k + \frac65$ & $\mA_3^{(4)}$ & $k \neq 3$ mod 5 \\[1mm]
    $E_{7,1} \otimes E_{7,1}$ & $-4k+14$ & $\frac12k + \frac34$ & $-k+\frac32$ & $\mA_3^{(1)}$ & \\[1mm]
    $A_{2,1} \otimes A_{2,1}$ & $16k+4$ & $\frac43k + \frac13$ & $\frac23k + \frac23$ & $\mB_8^{(1)}$ & $k \neq 2$ mod 3 \\[1mm]
    $A_{1,2}$ & $21k + \frac32$ & $\frac{13}{8}k + \frac{3}{16}$ & $k + \frac12$ & $\mathcal A_2^{(1)}$ & \\[1mm]
    $C_{2,1}$ & $19k + \frac52$ & $\frac{11}{8}k + \frac{5}{16}$ & $k + \frac12$ & $\mathcal A_2^{(1)}$ & \\[1mm]
    $A_{3,1}$ & $18k + 3$ & $\frac54k + \frac38$ & $k + \frac12$ & $\mB_3^{(4)}$ & \\[1mm]
    $B_{3,1}$ & $17k + \frac72$ & $\frac98k + \frac{7}{16}$ & $k + \frac12$ & $\mathcal A_2^{(1)}$ & \\[1mm]
    $A_{4,1}$ & $16k + 4$ & $\frac65k + \frac25$ & $\frac45k + \frac35$ & $\mB_4^{(1)}$  & $k \neq 3$ mod 5 \\[1mm]
    $B_{4,1}$ & $15k + \frac92$ & $\frac78k + \frac{9}{16}$ & $k + \frac12$ & $\mathcal A_2^{(1)}$ & \\[1mm]
    $D_{5,1}$ & $14k + 5$ & $\frac34k + \frac58$ & $k + \frac12$ & $\mB_3^{(4)}$ & \\[1mm]
    $B_{5,1}$ & $13k + \frac{11}{2}$ & $\frac58k + \frac{11}{16}$ & $k + \frac12$ & $\mathcal A_2^{(1)}$ & \\[1mm]
    $D_{6,1}$ & $12k + 6$ & $\frac12k + \frac34$ & $k + \frac12$ & $\mA_3^{(1)}$ & \\[1mm]
    $B_{6,1}$ & $11k + \frac{13}{2}$ & $\frac38k + \frac{13}{16}$ & $k + \frac12$ & $\mathcal A_2^{(1)}$ & \\[1mm]
    $D_{7,1}$ & $10k + 7$ & $\frac14k + \frac78$ & $k + \frac12$ & $\mB_3^{(4)}$ & \\[1mm]
    $B_{8,1}$ & $7k + \frac{17}{2}$ & $-\frac18k + \frac{17}{16}$ & $k + \frac12$ & $\mathcal A_2^{(1)}$ & \\[1mm]
    $D_{9,1}$ & $6k + 9$ & $-\frac14k + \frac98$ & $k + \frac12$ & $\mB_3^{(4)}$ & \\[1mm]
    $D_{10,1}$ & $4k + 10$ & $\frac14k + \frac78$ & $k + \frac12$ & $\mA_3^{(1)}$ & \\[1mm]
    \bottomrule
  \end{tabular}
  \caption{Quasi-character series from coset pairs}
  \label{qcseries}
\end{table}

Let us explain this table. The label for each row, appearing in the first column, is the minimal/WZW model with which we start the coset construction. This model corresponds to the $k=0$ case in its own line, while $k=1$ gives the dual coset with respect to a meromorphic $c=24$ CFT. We then extrapolated to all other values of $k$ (both positive and negative), and found that they indeed give quasi-characters. This is a conjecture that we have verified in large numbers of cases that are listed in Appendix \ref{verifying}. 

Going into more detail: the first seven entries actually did not appear in \cite{Gaberdiel:2016zke}. The first three of these were found later on in \cite{Hampapura:2016mmz} where the novel coset construction was generalised to the case of CFT without Kac-Moody symmetry. Such theories include minimal models as one would expect, and at least for three characters it is true that each model has a coset dual in the sense that a bilinear relation is satisfied.
The next four were missed in \cite{Gaberdiel:2016zke}. In these, the $k=0$ case is a direct-product CFT (this is also true of $\cM_{2,5}\otimes\cM_{2,5}$) but the coset, which is the $k=1$ entry, is not a direct-product theory . The remaining 14 entries in the table correspond to the 15 coset pairs in Table \ref{qcseries} of \cite{Gaberdiel:2016zke}. The reason why we appear to have one less entry is that the coset dual to $B_{8,1}$ is $E_{8,2}$, both are WZW models, and hence
the pair appeared twice in \cite{Gaberdiel:2016zke}.

As noted above, one can easily verify that the formula for $c,h_1,h_2$ fits a known CFT for $k=0$ and the coset dual for $k=1$ in every case. Indeed, this is the requirement on which our initial guess was based. If we now plug in the same formulae for the triple $(c,h_1,h_2)$ into the order-3 MLDE (with $\ell=0$) we find solutions with integral coefficients for every integer value of $k$. Moreover for $k\ne 0,1$ there are always some negative coefficients. Thus, these cases correspond to quasi-characters. 

We have previously noted that for any three-character CFT with $\ell=0$, the modular $\mS$-matrix can be computed using the technique in \cite{Mukhi:2019cpu}. This can immediately be extended to quasi-characters, since these satisfy the same type of MLDE as characters. One can assign fusion rules to quasi-characters by inserting their modular $\mc S$ transformation into the Verlinde formula. Table \ref{qcseries} identifies a fusion class for each line, and in the light of the previous remark this can be understood as the fusion class for the theories in the coset pair as well as of all quasi-characters in the same line. Note that different members of any given fusion class do not generally have the same $\mS$-matrix -- rather, $\mS$ varies periodically across each family of quasi-characters (as was already shown in \cite{Chandra:2018pjq} for order 2). However this does not happen for the $B_{r,1}$ and $D_{r,1}$ cases, where $\mS$ is actually the same for all members.

It is interesting that four of the fusion classes in Table \ref{qcseries}, namely $\cA_2^{(2)},\cB_3^{(1)},\cB_3^{(2)},\cB_3^{(3)}$ do not seem to be realised. There are no known CFT's with these fusion rules, but it is possible that there are as yet undiscovered quasi-characters with these fusion rules.

Let us now systematise and extend the results of Table \ref{qcseries}. First, we see that quite disparate models like the Ising model $\cM_{3,4}$ and the $A_{1,2}$ WZW model lie in the same fusion class as the $B_{r,1}$ models and in fact behave like special cases of those for $r=0,1$ (the latter is just the equivalence between SU(2)$_2$ and SO(3)$_1$). Accordingly we can subsume them under $B_{r,1}$ by allowing $r=0,1$. More important, we can extrapolate the above families to all the SO algebras, namely $B_{r,1}$ and $D_{r,1}$ for all positive $r$. For sufficiently large $r$, the theory at $k=0$ has $c>24$ and there is no longer any coset CFT in the family. Nevertheless, every case we have checked with such $r$ has confirmed the proposal that these are families of quasi-characters. Moreover, we find quasi-characters at {\em negative} values of $r$. However, the value $r=0$ is ruled out for the $D_{r,1}$ family.

In two of the families we have found additional quasi-characters by allowing $k$ to be a half-integer. However in the remaining examples there appear to be no quasi-characters at any fractional value of $k$. Thus our final proposal is that there exist (at least) the infinite set of families of quasi-characters listed in Table \ref{qcinfseries} for any $k$, where $k$ is an arbitrary integer except in two families where it can also be a half-integer as indicated in the Table.

A few caveats need to be made regarding the D-series entries in Table \ref{qcinfseries}. First of all, $r=4$ corresponds to a 2-character theory and should not be included in this Table. Second, $r=0$ is not allowed. Third, as indicated in the table, the case of $r=0$ mod 8 for $r\ne 0$ does not provide a valid family of quasi-characters, other than the $k=0,1$ cases which are known CFT for $r>0$.

\begin{table}[h!]
  \centering
  \begin{tabular}{cccccc}
    \toprule
    Fusion & \multirow{2}{*}{Theories} & \multirow{2}{*}{$c$} & \multirow{2}{*}{$h_1$} & \multirow{2}{*}{$h_2$} & \multirow{2}{*}{Remark} \\
    Class  & & & &  \\
    \midrule
    $\mA_2^{(3)}$ & $\mathcal M_{2,7}$ & $\frac{304}{7}k - \frac{68}{7}$ & $\frac{20}{7}k - \frac37$ & $\frac{18}{7}k - \frac27$ & $k \neq 4$ mod 7
    \\
    \midrule
    & $\mathcal M_{2,5} \otimes \mathcal M_{2,5}$ & $\frac{208}{5}k - \frac{44}{5}$ & $\frac{14}{5}k - \frac25$ & $\frac{12}{5}k-\frac15$ & $k \neq 3$ mod 5\\[0.4cm]
    $\mA_3^{(4)}$ & $G_{2,1} \otimes G_{2,1}$ & $\frac{64}{5}k + \frac{28}{5}$ & $\frac65k + \frac25$ & $\frac25k + \frac45$ & $k \neq 3$ mod 5 \\[0.4cm]
    & $F_{4,1} \otimes F_{4,1}$ & $\frac{16}{5}k + \frac{52}{5}$ & $\frac45k + \frac35$ & $-\frac25k + \frac65$ & $k \neq 3$ mod 5\\
    \midrule
    $\mB_4^{(1)}$ & $A_{4,1}$ & $16k + 4$ & $\frac65k + \frac25$ & $\frac45k + \frac35$ & $k\in\frac{\mathbb Z}{2}, 2k \neq 1$ mod 5 \\
    \midrule
    $\mA_2^{(1)}$ & $B_{r,1}$ & $(23-2r)k + \frac{2r+1}{2}$ & $\frac{(15-2r)}{8}k + \frac{2r+1}{16}$ & $k+\frac12$ & $r=0,1$ incl. \\
    \midrule
    \multirow{2}{*}{$\mA_3^{(1)}$}& $E_{7,1} \otimes E_{7,1}$ & $-4k+14$ & $\frac12k + \frac34$ & $-k+\frac32$ & \\[0.2cm]
    & $D_{r,1}$ & $2(12-r)k + r$ & $\frac{(8-r)}{4}k + \frac{r}{8}$ & $k + \frac12$ & $r = 2$ mod 4 \\
    \midrule
    $\mB_3^{(4)}$ & $D_{r,1}$ & $2(12-r)k + r$ & $\frac{(8-r)}{4}k + \frac{r}{8}$ & $k + \frac12$ & $\begin{matrix} r \neq 2~ {\rm mod}~ 4\\  r\ne 0 ~{\rm mod}~ 8\end{matrix}$ \\
    \midrule
    $\mB_8^{(1)}$ & $A_{2,1} \otimes A_{2,1}$ & $16k+4$ & $\frac43k + \frac13$ & $\frac23k + \frac23$ & $k\in\frac{\mathbb Z}{2}, 2k\ne 1$ mod 3 \\
    \bottomrule
  \end{tabular}
  \caption{Infinitely many sets of quasi-character families for each fusion class. Here $n,r\in\mathbb Z$ subject to the restrictions above. }
  \label{qcinfseries}
\end{table}

In the case of order 2, we were able to restrict the quasi-character series \eref{c.quasi} to $k\ge 0$ by suitably exchanging characters. However the corresponding process is not so straightforward here. Hence we keep both positive and negative values of $k$. Moreover as noted above, even for $r$ negative one finds quasi-characters though in this case there are no admissible characters in the family.

The values of $c,h_1,h_2$ for the $B_{r,1}$ and $D_{r,1}$ families can easily be combined into a single family parametrised by the dimension $d$ of the fundamental of $B_r,D_r$, which is $2r+1,2r$ respectively. However since $B_{r,1}$ and $D_{r,1}$ have different fusion rules (and in fact $D_{r,1}$ also splits further into two fusion classes) we prefer to list them separately.

Let us make a comment about the ``presentation'' of the characters. It was noted in Footnote \ref{unpres} that characters arising from an MLDE can be considered in multiple presentations, by picking any one of the characters to correspond to the identity and identifying its exponent with $-\frac{c}{24}$. Among these, there is always a distinguished unitary presentation for which the most negative exponent corresponds to the identity character. This was the choice we made in the two-character case. However, in the present situation we have made a different choice. Because both $r$ and $k$ run over positive and negative values, it is unnatural to keep changing the presentations within a given family. Instead we fix a presentation such that the $r>0,k=0$ theories are exhibited in their familiar forms as minimal or WZW models. For other values of $r,k$ the presentation may be either unitary or non-unitary. 

It turns out that, like admissible characters, quasi-characters too satisfy a novel coset relation as originally defined in \cite{Gaberdiel:2016zke}, or more precisely the bilinear relation associated to it. To see this, note first that every entry in Table \ref{qcseries} has a central charge of the form $Ak+B$ for some constants $A$ and $B$. Let us denote by $\chi_i^{(A,B,k)}$ the corresponding quasi-characters. By construction, each admissible character $\chi_i^{(A,B,k=0)}$, corresponding to a genuine CFT, has a coset pair $\chi_i^{(A,B,k=1)}$ in its family. Hence they obey a holomorphic bilinear relation with respect to a $c=24$ meromorphic CFT:
\be
\sum_{i=0}^2 \chi_i^{(A,B,k=0)}(\tau)\,\chi_i^{(A,B,k=1)}(\tau)=j(\tau)+\cN
\ee
for some integer $\cN$, where $j(\tau)$ is the Klein invariant. Here $\cN$ should correspond to one of the theories in \cite{Schellekens:1992db}.

Now we may ask whether there is a bilinear relation satisfied by pairs of quasi-characters in Table \ref{qcinfseries} whose $c$ values add up to 24. To address this we observe that in every row of Table \ref{qcinfseries}, parametrised by $c=Ak+B$, one has $A+2B=24$. Hence if we consider $\chi_i^{(A,B,k)}$ and $\chi_i^{(A,B,1-k)}$ for any $k$, this pair could also potentially satisfy a bilinear relation of the same form:
\be
\sum_{i=0}^2 \chi_i^{(A,B,k)}(\tau)\,\chi_i^{(A,B,1-k)}(\tau)=j(\tau)+\cN
\ee
for some $\cN$, not necessarily integer. By construction the LHS has a $q$-series that starts with $q^{-1}$, just like the proposed RHS. Now one can test whether the two sides match by comparing large numbers of terms in the $q$-series of each factor on the LHS, case by case as $k$ varies. We have carried this out for the cases listed in Appendix \ref{verifying}.

The result is that the coset relation does hold, but with one caveat. Recall that nontrivial ground-state degeneracies $D_i$ generically had to be included to ensure integrality of the $q$-expansion. For admissible characters the $D_i$  are of course positive. However for quasi-characters there is no a priori rule to determine the sign of $D_i$. Modular invariance of the ``partition function'' for quasi-characters does not help because the $D_i$ are squared in that expression. However in the bilinear relation one finds a nontrivial condition on the sign of the product $D_iD'_i$, of the ground-state degeneracies of each of the coset pairs. Only if the correct sign is chosen (and this sometimes comes out to be negative) does the coset relation hold. In summary, quasi-character pairs have been verified to obey the coset relation to high orders in the $q$-expansion for some suitable choice of the sign of $D_iD'_i$. This is quite a miraculous fact, going far beyond the original proposal of \cite{Gaberdiel:2016zke} which dealt only with admissible characters\footnote{The analogous phenomenon has been previously noted in order 2 \cite{Hampapura:2016mmz,Chandra:2018pjq}.}. It will play a significant role in the considerations of the following sections.

In Appendix \ref{verifying} we have summarised the range of values of $r,k$ and the order in the $q$-series for which we have verified the conjectured quasi-character families in Table \ref{qcinfseries}. We also summarise cases and the order in $q$ to which we have verified the bilinear relation.

\subsection{Modular properties of quasi-character series}
\label{modS}

In this subsection we discuss the modular properties of the quasi-character families we have proposed in Table \ref{qcinfseries}. We have studied each of these families and calculated the modular $\mS$-matrix using techniques from \cite{Mukhi:2019cpu}. The relevant result from these references is a general formula for the  characters in order 3 with $\ell=0$ (which applies equally well to admissible characters or quasi-characters). These are  given in terms of the function:
\be
\lambda(\tau)\equiv \frac{\theta_2^4(\tau)}{\theta_3^4(\tau)}
\ee
by the following two-variable contour integrals:
\be
\begin{split}
\chi_0 &= N_0\big(\lambda(1-\lambda)\big)^\alpha\int_1^\infty dt_2 \int_1^\infty dt_1 
\big[t_1 (t_1-1)(t_1-\lambda)\big]^a \big[t_2 (t_2-1)(t_2-\lambda)\big]^a (t_2-t_1)^{2\rho}\\
\chi_1 &= N_1\big(\lambda(1-\lambda)\big)^\alpha\int_1^\infty dt_2 \int_0^\lambda dt_1 
\big[t_1 (1-t_1)(\lambda -t_1)\big]^a \big[t_2 (t_2-1)(t_2-\lambda)\big]^a (t_2-t_1)^{2\rho}\\
\chi_2 &= N_2\big(\lambda(1-\lambda)\big)^\alpha\int_0^\lambda dt_2 \int_0^\lambda dt_1 
\big[t_1 (1-t_1)(\lambda -t_1)\big]^a \big[t_2 (1-t_2)(\lambda-t_2)\big]^a (t_2-t_1)^{2\rho}
\end{split}
\ee
where $\alpha=-2(\frac{\rho+1}{3}+a)$. The parameters $a,\rho$ are related to the central charge and conformal dimensions as:
\be
  c = 8(3a+\rho+1) , \quad h_1 = a+\frac12 , \quad h_2 = 2a+\rho+1
\ee

In terms of the $\lambda$ variable, the modular $\mS$-transformation is $\lambda\to 1-\lambda$. As originally shown in \cite{Dotsenko:1984ad,Dotsenko:1984nm}, under this transformation the above integrals come back to calculable linear combinations of themselves. However this is not sufficient to determine the modular-transformation matrix. One needs to know the degeneracies $D_i$ and normalisations $N_i,i=0,1,2$  in order to do this. These were determined in \cite{Mathur:1988gt,Mukhi:1989qk,Mukhi:2019cpu}, the last of which provided  an elegant recursive algorithm to compute $\mS$.

Recalling the discussion on the sign of ground-state degeneracies $D_i$ of the previous subsection, we now note that this in turn determines the signs of entries in the $\mS$-matrix. Thus we should specify the sign of $D_i$ when quoting the $\mS$-matrix for the members of each family. In the following this sign may be assumed positive except where explicitly noted as negative. As an example, for the $A_{4,1}$ family we have: 
\be
\begin{split}
  A_{4,1} \text{ family} : \quad
  \mS&=\begin{pmatrix}
    \frac{1}{\sqrt5} & \frac{2}{\sqrt5} &  \frac{2}{\sqrt5} \\[2mm]
    \frac{1}{\sqrt5} & \frac{5-\sqrt5}{10} &   \frac{-5-\sqrt5}{10}\\[2mm]
    \frac{1}{\sqrt5} &  \frac{-5-\sqrt5}{10} &  \frac{5-\sqrt5}{10} \\
  \end{pmatrix} \quad
  \begin{matrix}
    k=0 &\\
    2k=2,5~{\rm mod}~10 & \\ 2k=7,10 ~{\rm mod}~10& \; (D_2<0)
  \end{matrix} \\[0.2cm]
  \mS&=\begin{pmatrix}
    -\frac{1}{\sqrt5} & \frac{2}{\sqrt5} &  \frac{2}{\sqrt5} \\[2mm]
    \frac{1}{\sqrt5} & \frac{5+\sqrt5}{10} &   \frac{-5+\sqrt5}{10}\\[2mm]
    \frac{1}{\sqrt5} &  \frac{-5+\sqrt5}{10} &  \frac{5+\sqrt5}{10} \\
  \end{pmatrix} \quad
  \begin{matrix}
    2k=3,4~{\rm mod}~10 & \\ 2k=8,9~{\rm mod}~10 & \;\; (D_2 <0)
  \end{matrix}
\end{split}
\label{A41S}
\ee

Two more examples that cover a large number of families are:
\be
D_{r,1}~\hbox{family, all }r:\qquad
\mS=\begin{pmatrix}
    \frac12 & 1 & \frac12 \\
    \frac12 & 0 & -\frac12 \\
    \frac12 & -1 & \frac12 \\
  \end{pmatrix}
\ee
and:
\be
B_{r,1}~\hbox{family, all }r:\qquad
\mS=\begin{pmatrix}
    \frac12 & \frac{1}{\sqrt{2}} & \frac12 \\[0.2cm]
    \frac{1}{\sqrt{2}} & 0 & -\frac{1}{\sqrt{2}} \\[0.2cm]
    \frac12 & -\frac{1}{\sqrt{2}} & \frac12 \\
  \end{pmatrix}
\ee
Here we find an $r$-dependent pattern for the signs of $D_1,D_2$ in order to have the above $\mS$-matrix with the given signs. For example, in the $D_{3,1}$ family  one finds  positive ground-state degeneracies for $k=0,1,2,3,4$ and negative $D_2$ for $k=5,6,7,8$. Similar patterns arise in other cases, but they are somewhat complicated and we do not list them here.

The $\mS$-matrices for the remaining quasi-character families can be computed using the algorithm presented in 
\cite{Mukhi:2019cpu}.

\section{Admissible characters in order 3}

\label{three.admiss}

\subsection{Adding quasi-characters}

\label{adding}

We now provide some general results about how to generate new quasi-characters with $\ell>0$ by adding two quasi-characters with $\ell=0$. Consider two sets of (quasi-)characters with central charges $c,c'$ and conformal dimensions $h_i,h_i'$. They have the general form:
\be
\begin{split}
  \chi_i &= q^{\alpha_i}\left(a_0^{(i)} + a_1^{(i)} q + a_2^{(i)} q^2 + \cdots\right)\\
  \chi'_i &= q^{\alpha'_i}\left({a'_0}^{(i)} + {a'_1}^{(i)} q + {a'_2}^{(i)} q^2 + \cdots\right)
\end{split}
\ee
where the exponents $\alpha_i,\alpha'_i$ are related in the usual way to the central charge and dimensions. The $a_n,a'_n$ are integers, but not necessarily positive.

Our goal is to add $\chi_i'$ to $\chi_i$, with a relative coefficient, to generate new objects of the form:
\be
 {\wtd \chi_i} \equiv \chi'_i+ N \chi_i= q^{{\wtd\alpha}_i}\left({\wtd a}_0^{(i)} + {\wtd a}_1^{(i)} q + {\wtd a}_2^{(i)} q^2 + \cdots\right)
\ee
where the sum satisfies two conditions: (i) it must have a well-defined modular $\mS$-matrix, (ii) it must have a series expansion in integer powers of $q$ as indicated above. In a later sub-section we will use such sums to generate admissible characters.

The first requirement tells us that $\chi_i,\chi'_i$ should lie in the same quasi-character family and moreover their modular $\mS$-matrix must be the same. Since every quasi-character family has central charges (in the presentation of Table \ref{qcinfseries}) of the form:
\be
c=Ak+B
\ee
for all integers $k$ and some rational numbers $A,B$, we see that:
\be
c'-c=(k'-k)A
\ee
The second requirement implies that $c'-c$ must be an integer multiple of 24:
\be
c'-c=24m
\label{cdiff}
\ee
Hence we must ensure that $(k'-k)A$ is an integral multiple of 24.

Let us now consider a few examples. Here we are going to quote the $q$-expansion only to low orders in order not to clutter the equations, however any general conclusions we have drawn about their behaviour are based on examining the $q$-series to an order of around 1000, as listed in Appendix \ref{verifying}. The conclusions are still conjectural of course.

\subsubsection*{Example: $A_{4,1}$ series: $k=1,\frac52$}

From Table \ref{qcinfseries}, this series has $c=16k+4, h_1=\frac65 k+\frac25, h_2=\frac45 k+\frac35$ where $k$ is an integer or half-integer. For $\chi_i$ we take $k=1$ because this gives an admissible character. From the Table we read off that $c=20,h_1=\frac85,h_2=\frac75$. To this we will add $\chi'_i$, a set of quasi-characters with $k'=\frac{5}{2}$, for which $c'=44,h'_1=\frac{17}{5},h'_2=\frac{13}{5}$. From \eref{A41S} we see that both $k=1$ and $k'=\frac52$ have the same $\mS$:
\be
\mS=\begin{pmatrix}
    \frac{1}{\sqrt5} & \frac{2}{\sqrt5} &  \frac{2}{\sqrt5} \\[2mm]
    \frac{1}{\sqrt5} & \frac{5-\sqrt5}{10} &   \frac{-5-\sqrt5}{10}\\[2mm]
    \frac{1}{\sqrt5} &  \frac{-5-\sqrt5}{10} &  \frac{5-\sqrt5}{10} \\
  \end{pmatrix}
\ee
The same equation also tells us that the ground-state degeneracies for $k'=\frac52$ must be positive. Now we can add them.

The characters have the expansions:
\be
\begin{split}
c=20: &\\
\chi_0 &=q^{-\frac56}\left(1 + 120 q + 62630 q^2 + 3562760 q^3 +\cdots\right)\\
\chi_1 &= q^\frac{23}{30}\left(8125 + 805000 q + 26708750 q^2 + 
 512587500 q^3+\cdots\right)\\
\chi_2 &=  q^\frac{17}{30}\left(2500 + 361250 q + 13985000 q^2+ 
 293167500 q^3 +\cdots\right)\\
c=44: &\\
\chi'_0 &= q^{-\frac{11}{6}}\left(4 - 319q - 32824 q^2 + 103089030 q^3+\cdots\right)\\
\chi'_1 &= q^{\frac{23}{30}}\Big(2578125 + 6008750000 q + 1314955468750 q^2 \\
&\qquad\qquad + 
 117965856250000 q^3 + \cdots\Big)\\
\chi'_2 &= q^{\frac{47}{30}}\Big(1717031250 + 489093750000 q +  50494582187500 q^2\\
& \qquad\qquad + 2881407247812500 q^3 +\cdots\Big)
\end{split}
\ee
An amusing point is that we have been forced to flip the definition of $h'_1$ and $h'_2$ in order that the characters can be added, as one can see by comparing the exponents of $\chi_i,\chi'_i$ above. This is a basis change and should have been dictated by the $\mS$-matrix. However in this particular example the $\mS$-matrix (see \eref{A41S}) is actually invariant under this flip and therefore allows either ordering. The right choice is then the one for which adding quasi-characters yields a result with a sensible $q$-expansion.

Now defining $\tchi_i=\chi'_i+N\chi_i$, we get:
\be
\begin{split}
\tchi_0 &=q^{-\frac{11}{6}} (4 + (-319 + N) q + 8 (-4103 + 15 N) q^2 + 10 (10308903 + 6263 N) q^3 +\cdots\\
\tchi_1 &=q^\frac{23}{30}\Big((2578125 + 8125 N)  + (6008750000 + 805000 N) q + (1314955468750 + 26708750 N) q^2\\
&\qquad\qquad + (117965856250000 + 512587500 N) q^3+\cdots)
\\
\tchi_2 &=q^\frac{17}{30}\Big(2500 N  + (1717031250 + 361250 N) q +  (489093750000 + 13985000 N) q^2\\
&\qquad\qquad +  (50494582187500 + 293167500 N)q^3+\cdots \Big)
\end{split}
\label{20.44}
\ee
As expected, the sum has a well-defined $\mS$-matrix and a sensible $q$-series. Its exponents are $\tc=44,\tilh_1=\frac{13}{5},\tilh_2=\frac{12}{5}$ and from this one finds $\ell=6$.

\subsubsection*{Example: $D_{r,1}$ series: $(r,k)=(5,0),(-5,1)$}

The next example is taken from the $D_{r,1}$ families in Table \ref{qcinfseries}. A general class of examples can be found within this family by taking the pair $(r,k=0)$ and $(-r,k=1)$ for any $r$. It is easy to verify that for all such pairs, $c'-c=24$. Similarly, in the $B_{r,1}$ series one may take $(r,k=0)$ and $(-r-1,k=1)$ and again $c'-c=24$. 

Let us now take $\chi_i$ to correspond to $(r,k)=(5,0)$. This is an admissible character. From the Table we read off that $c=5,h_1=\frac58,h_2=\half$. To this, we add a quasi-character $\chi'_i$ with $c'=29$. This can be achieved by the choice $(r',k')=(-5,1)$ which leads to $c'=29,h'_1=\frac{21}{8},h'_2=\frac32$. 

From Subsection \ref{modS} we see that both cases have 
the same modular $\mS$-matrix, namely:
\be
\mS=\begin{pmatrix}
    \frac12 & 1 & \frac12 \\[0.2cm]
    \frac12 & 0 & -\frac12 \\[0.2cm]
    \frac12 & -1 & \frac12 \\
  \end{pmatrix}
\ee
for a suitable choice of sign of the degeneracies. With this choice of signs, the $q$-series for the two sets are:
\be
\begin{split}
c=5: &\\
\chi_0&=q^{-\frac{5}{24}}(1+45q+310q^2+1555q^3+\cdots)\\
\chi_1&=q^{\frac{5}{12}}(16+160q+880q^2+3680q^3+\cdots)\\
\chi_2&=q^{\frac{7}{24}}(10+130q+712q^2+3130q^3+\cdots)\\
c'=29: &\\
\chi'_0 & = q^{-\frac{29}{24}}(1-319q+78590q^2+25022911q^3+\cdots)\\
\chi'_1 & = q^{\frac{17}{12}}(3801088+397672448q+17830903808q^2+486562070528q^3+\cdots)\\
\chi'_2 &=q^{\frac{7}{24}}(-1624+1921192q+235569088q^2+11440410216q^3+\cdots)
\end{split}
\ee
If we now define $\tchi_i=\chi'_i+N\chi_i$ where $N$ is  an arbitrary integer, we get:
\be
\begin{split}
\tchi_0 &=
q^{-\frac{29}{24}}\left(1+(-319+N)q+5(9N+15718)q^2
+(310N+25022911)q^3+\cdots\right)\\
\tchi_1 &=
q^{\frac{5}{12}}\Big(16 N +(160 N+3801088) q+(880N+397672448)q^2\\
&\qquad\quad  +(3680 N+17830903808)q^3+\cdots\Big)\\
\tchi_2 &=
q^{\frac{7}{24}}\Big((10 N-1624) +(130 N+1921192) q+(712N+235569088)q^2\\
&\qquad\quad  +(3130 N+11440410216)q^3+\cdots\Big)
\end{split}
\label{5.29}
\ee
This sum again has a well-defined modular $\mS$-matrix and sensible $q$-series. Its exponents are $\tc=29,\tilh_1=\frac{13}{8},\tilh_2=\frac32$ from which one can verify that $\ell=6$. 

\subsubsection*{Example: $D_{r,1}$ series: $(r,k)=(-5,0),(5,1)$}

This is an amusing example because it is obtained from the previous one by the replacement $r\to -r$ keeping the $k$ values fixed, but its behaviour is rather different. The $\mS$-matrix is actually the same as in the previous example. Now $(r,k)=(-5,0)$ has $c=-5$ while $(r',k')=(5,1)$ with $c'=19$. This time the second one is an admissible character (it is the coset of $(5,0)$ with respect to a meromorphic $c=24$ theory). The sum is found to be:
\be
\begin{split}
\tchi_0 &= q^{-\frac{19}{24}}\left(1 + (171+64 N) q + (60895 + 
 3520 N) q^2 + (2958376 + 52160 N)q^3+\cdots\right)\\
\tchi_1 &=
q^{-\frac{5}{12}}\Big(N + (2432 - 10 N)q + (296192 + 
 45 N) q^2 + (10216832 - 130 N) q^3 +\cdots\Big)\\
\tchi_2 &=
q^{\frac{17}{24}}\Big((5016 - 640 N)  + (483208 - 
 14720 N) q + (15160024 - 163968 N) q^2\\
& \qquad\quad + 
 (275040656 - 1260800 N) q^3 +\cdots\Big)
\end{split}
\label{-5.19}
\ee
We notice that, unlike the previous example, there are oscillating signs in the last character. We will return to this point below.

\subsubsection*{Example: $A_{2,1} \otimes A_{2,1}$ series: $k=0,\frac32$}

The last example we will consider is from the $A_{2,1} \otimes A_{2,1}$ series which has $c=16k+4, h_1=\frac43 k+\frac13, h_2=\frac23 k+\frac23$ where $k$ is an integer or half-integer. For $\chi_i$ we will take $k=0$ because this gives an admissible character, that of the $A_{2,1} \otimes A_{2,1}$ WZW model. From the Table \ref{qcinfseries}, we read off that $c=4, h_1=\frac13, h_2=\frac23$. To this we add $\chi_i'$ which corresponds to $k'=\frac32$, for which $c'=28, h'_1=\frac73, h'_2=\frac53$. Both these characters have the same $\mS$ matrix which is given by:
\be
\mS = \begin{pmatrix}
  \frac13 & \frac43 & \frac43 \\[0.2cm]
  \frac13 & \frac13 & -\frac23 \\[0.2cm]
  \frac13 & -\frac23 & \frac13 \\
\end{pmatrix}
\ee
This allows us to take a linear combination of $\chi_i$ and $\chi_i'$ by defining $\tchi_i=\chi'_i+N\chi_i$, which has the following $q$-expansion:
\be
\begin{split}
  \tchi_0 &= q^{-\frac76} \Big(1 + (-77+ N)q + (64274+16 N)q^2 + (14583702+98 N)q^3+ \cdots \Big) \\
  \tchi_1 &= q^{\frac16}\Big(3 N + (492075 + 33 N)q + (63930384+150 N)q^2 \\
  & \qquad \;\; + (3137548932+564 N)q^3 + \cdots \Big) \\
  \tchi_2 &= q^{\frac12}\Big((5103 + 9 N) + (2924019 + 54 N)q + (253103697 + 243 N)q^2 \\
  & \qquad \;\; + (10060647606 + 828 N)q^3 + \cdots \Big)
\end{split}
\label{4.28}
\ee

In each of the above examples, $\tchi_i$ has Wronskian index $\ell=6$, as one can verify using the Riemann-Roch theorem. In the next subsection we show how to determine $\ell$ in general when adding a pair of quasi-characters. We also note that for general values of $N$ the $\tchi_i$ are generically quasi-characters, but we will show later on that for some ranges of $N$ they correspond to admissible characters.

\subsection{Wronskian index of a sum of two quasi-characters}

Here we compute the Wronskian index  $\ell$ corresponding to the sum of a pair of quasi-characters whose central charges differ by $24m$. Without loss of generality we have assumed that $c' > c$ and hence $m$ takes values over positive integers only. We now want to find out the values of the exponents $\wtd\alpha_i$ for the sum of the two original quasi-characters.

For the identity character we see that since $c' > c$, $-\frac{c'}{24} < -\frac{c}{24}$ and hence $\chi_0'$ is more singular than $\chi_0$. Therefore the leading term of $\chi_0$ adds to the $m^{th}$ term of $\chi_0'$, and we have
\be
  {\wtd\alpha}_0 = \alpha_0' = -\frac{c'}{24}
\ee
Next, consider $\wtd\alpha_1$. We can write $\alpha_1$ and $\alpha_1'$ in the following way
\be
  \alpha_1 = -\frac{c}{24} + h_1 = -\frac{c'}{24} + h_1 + m, \qquad \quad \alpha_1' = -\frac{c'}{24} + h_1'
\ee
The exponent $\wtd\alpha_1$ corresponds to the more singular of the above two exponents, hence we have:
\be
  \wtd\alpha_1 = \wtd\alpha_0 + \min(h_1',h_1+m)
\ee
By the same reasoning we see that:
\be
\wtd\alpha_2 = \wtd\alpha_0 + \min(h_2',h_2+m)
\ee
Depending on which values are smaller in the above, there are four possibilities.

Consider the case where $h_1' < h_1 + m$. Combining \eref{cdiff} with the Riemann-Roch relation \eref{Riemann.Roch.3}, we find that:
\be
h_1' - h_1 + h_2' - h_2 = \frac{c'}{8} - \frac{c}{8} = 3m 
\ee
This in turn implies:
\be
h_1' - h_1 - m = 2m + h_2 - h_2' < 0
\ee
from which we conclude that:
\be
h_2' > h_2 + 2m
\ee
Thus we see that both $h_1'$ and $h_2'$ cannot be less than $h_1 + m$ and $h_2 + m$ simultaneously. So we can restrict ourselves to the remaining three cases.

Let us start with the case where $h_i + m < h_i'$ for both $i=1,2$. Then the $\wtd\alpha_i$ are:
\be
  \wtd\alpha_0 = -\frac{c'}{24}, \quad \wtd\alpha_1 = -\frac{c'}{24} + h_1 + m, \quad \wtd\alpha_2 = -\frac{c'}{24} + h_2 + m
\label{alphatilde}
\ee
We now use the Riemann-Roch relation again to find the value of the Wronskian index $\ell$ for this set of exponents:
\be
\sum_{i=0}^2 \wtd\alpha_i = \frac12 - \frac{\ell}{6} 
\ee
Inverting this and using Eqs.(\ref{alphatilde}, \ref{cdiff}, \ref{Riemann.Roch.3}) we find:
\be
\ell = 6m
\ee
Repeating this for the other two cases, we get the $\ell$ values listed in Table \ref{ellvaluetable}.

\begin{table}[h!]
  \centering
  \begin{tabular}{ccc}
    \toprule
    $\min(h_1',h_1+m)$ & $\min(h_2',h_2+m)$ & $\ell$ \\
    \midrule
    $h_1'$ & $h_2+m$ & $6(h_2'-h_2-m)$ \\[2mm]
    $h_1+m$ & $h_2'$ & $6(h_1'-h_1-m)$ \\[2mm]
    $h_1+m$ & $h_2+m$ & $6m$ \\[2mm]
    \bottomrule
  \end{tabular}
  \caption{Possible values of the Wronskian index $\ell$}
  \label{ellvaluetable}
\end{table}

The above construction can be generalised to the sums of three or more quasi-characters. The arguments are very similar in spirit -- each of the summands has to have the same central charge mod 24, and also the same modular $\mS$-matrix. One can easily verify that the sums obtained in this way have larger and larger values of $\ell$.

\subsection{New admissible characters with $\ell>0$}

In the previous subsection we showed how to add sets of quasi-characters with $\ell=0$ in such a way that the sum is  a set of quasi-characters with $\ell>0$. We found that the sum has values of $\ell$ that are multiples of 6. The resulting quasi-characters depend on arbitrary integers, the coefficients of each summand. Now we look for examples where the arbitrary integers can be chosen in such a way that all the negative coefficients are removed. This will impose bounds on the integers. 

First, let us classify the type of behaviour of quasi-characters with respect to the occurrence of minus signs. In order 2 this was done in \cite{Chandra:2018pjq} and, as we discussed above, the result was that there are two types of quasi-characters, which were labelled Type I and Type II depending on the behaviour of the identity character.

In order 3, the situation is more complex. We classify {\em each} of the three quasi-characters into one of the following classes:
\begin{itemize}
\item
Class A: The $q$-expansion has only non-negative coefficients.

\item
Class B: After a finite power of the $q$ expansion, there are only positive coefficients (assuming the ground state degeneracy is chosen positive).

\item
Class C: After a finite power of the $q$ expansion there are only negative coefficients (again assuming the ground state degeneracy is chosen positive).

\item 
Class D: The asymptotic sign of the coefficients does not stabilize but oscillates between positive and negative values.

\end{itemize}

Note that each of the A,B,C behaviours is defined in a convention where the first term is chosen positive, corresponding to a positive degeneracy $D_i$. If instead one chooses $D_i<0$, which is sometimes required, then the behaviour flips in an obvious way. 

One has to specify one of A,B,C,D for each of the three $\chi_i$, so the class is specified by a triple. In this notation, admissible characters correspond to Class AAA. We have found examples of type AAD, ABD, ACD, BBB, BBC and many more. Clearly the variety of behaviours is far more complicated than in order 2. Going back to the examples in Subsection \ref{adding}, recall that one of the two sets of quasi-characters in each example is admissible or AAA. We have verified to high orders ($\sim q^{5000}$) that in the first and last examples $\chi_i'$ is BAA, in the second example $\chi_i'$ is BAC and in the third example $\chi_i$ is ADA.

Now the idea is to generate admissible characters by suitably choosing the coefficient $N$. Let us first examine this for each of the four examples of Subsection \ref{adding}. In the first one, involving a pair with $(c,c')=(20,44)$, the sum in \eref{20.44} becomes admissible for $N\ge 319$. In the second example, involving a pair with $(c,c')=(5,29)$, the sum in \eref{5.29} also becomes admissible for $N\ge 319$ (this is just a coincidence as far as we can see). In the third example, involving a pair with $(c,c')=(-5,19)$, we find that the character appears to become admissible for $-2 \le N\le 7$. This case is more tricky because of the type D behaviour of the middle quasi-character. Still, putting $N$ equal to $-2$ and 7 by turns, we have verified that the result appears admissible up to order $q^{5000}$. Finally in the last example, involving a pair with $(c,c')=(4,28)$, we find the characters become admissible for $N \geq 77$. Clearly many more examples can be generated in this way, though we do not yet have a general theory or classification.

\subsection{Cosets and three-character CFT with $\ell\ge 6$}

In this subsection we present an example of a coset CFT that matches with one of the admissible characters constructed in subsection \ref{adding}. This will provide a demonstration that at least some of the admissible characters obtained by adding quasi-characters coincide, for a specific value of the coefficient $N$, with those of a genuine conformal field theory.

The novel coset construction of \cite{Gaberdiel:2016zke} tells us that if we coset a meromorphic CFT of central charge $8N$ by a CFT with $p$ characters and Wronskian index $\ell$, the coset theory has Wronskian index:
\be
\tilde\ell=p^2+(2N-1)p-6(n_1+n_2)-\ell
\ee
Here $n_1,n_2$ are integers such that $h_i+\tilh_i=n_i,i=1,2$. It has been argued in the above reference that the minimal value of $n_1,n_2$ in such examples is 2. If we assume these values, and consider a three-character $\ell=0$ theory in the denominator of the coset, the above formula reduces to:
\be
\tilde\ell=6N-18
\ee
Thus if we start with a 32-dimensional meromorphic CFT based on an even self-dual lattice with a complete root system (for which $N=4$) the coset theory has $\tilde\ell=6$. The coset construction defines a genuine CFT. One can compute
the characters of this CFT and try to match them with one or more of the admissible characters previously constructed. 

As an example we consider the CFT of 32 free bosons on the Kervaire lattice \cite{Kervaire:1994} with complete root system $A_1^8A_3^8$ \cite{King:2001}. The  corresponding level-1 Kac-Moody algebra has dimension 144 and the single character of this theory is:
\be
\chi=j^\frac13(j-848)=q^{-\frac43}(1+144q+\cdots)
\ee 
We now take the quotient of this meromorphic CFT by the WZW model $A_{3,1}$. The result is a three-character CFT with $c=29$, $h_1=\frac{13}{8}$, $h_2=\frac32$. Its Kac-Moody algebra has dimension 129. 

We can now compare this with the family of admissible characters in our example \eref{5.29}. That entire family has precisely $c=29$, $h_1=\frac{13}{8}$, $h_2=\frac32$ for any non-zero value of $N$. The last step is to note that the dimension of the would-be Kac-Moody algebra there is $N-319$. Setting $N=448$, the coefficient becomes 129. Thus we have reproduced a set of $\ell=6$ admissible characters with the right exponents and dimension of Kac-Moody algebra to correspond to our coset theory above. To verify that this is indeed the case, we evaluate the bilinear product of the corresponding admissible characters with the $A_{3,1}$ WZW characters. The result turns out to be $j^\frac13(j-848)$, which is exactly the character of the original meromorphic theory.

As another example, we consider the CFT of 32 free bosons on the Kervaire lattice with complete root sysytem $A_2^2 A_{14}^2$. The corresponding level-1 Kac-Moody algebra has dimension 448 and the single character of this theory is:
\be
  \chi = j^{\frac13} (j-544) = q^{-\frac43} (1+448q+\cdots)
\ee
We now take the quotient of this meromorphic CFT by the $A_{2,1} \otimes A_{2,1}$ WZW model. The result is a three-character CFT with $c=28, h_1=\frac53, h_2=\frac43$. Its Kac-Moody algebra has dimension 432.

We can now compare this with the family of admissible characters in our example \eref{4.28}. The entire family has $c=28, h_1=\frac43, h_2=\frac53$ for any non-zero value of $N$ and the dimension of the would-be Kac-Moody algebra is $N-77$. Thus, setting $N=509$, we reproduce a set of $\ell=6$ admissible characters with the right exponents and dimension of Kac-Moody algebra, after switching $\tilde h_1$ and $\tilde h_2$ to match the coset theory above. Once again, to verify that this indeed the case, we evaluate the bilinear product of the corresponding admissible characters with the $A_{2,1} \otimes A_{2,1}$ WZW characters. The result turns out to be $j^{\frac13} (j-544)$, which is exactly the character of the original meromorphic CFT.

So far we have not been able to find coset theories to correspond to our examples in \eref{20.44} and \eref{-5.19}, but we also cannot rule out the existence of such theories. 

One outstanding puzzle is that we have not found any set of quasi- or admissible characters in order 3 having $\ell=1,2,3,4,5$ mod 6. For two-character theories one can prove that odd values are forbidden \cite{Naculich:1988xv}, and the classification of \cite{Chandra:2018pjq} includes all even values. It may be the case that in order 3, the missing values of $\ell$ are all forbidden for some reason, or else they exist and require an entirely new construction. More work is required to settle this issue.

\section{Discussion and conclusions}

We have proposed a large collection of families of quasi-characters with vanishing Wronskian index ($\ell=0$) and provided evidence for their existence. This provides at least a partial answer to the fascinating mathematical question of where in the parameter space of vector-valued modular functions in order 3 one finds a $q$-expansion with integral coefficients. In addition, an analysis of the modular $\mS$-matrix of our quasi-characters based on the contour-integral proposal of \cite{Mukhi:1989qk, Mukhi:2019cpu} allows us to determine when two or more quasi-characters can be added to each other preserving modularity. The sums so obtained have Wronskian index $\ell=6m$ for some positive integer $m$. We worked out a few examples where the sum has $\ell=6$. 

Next we showed that among these sums, one can find admissible characters where the $q$-expansion is completely non-negative. These can potentially correspond to CFT. Finally we identified a couple of novel coset models having three characters and $\ell=6$ and identified them as special cases of our admissible characters. This shows that the set of admissible characters generated by our construction that correspond to actual CFT with $\ell=6$ is non-empty.

Our method involved two types of computations: an evaluation of the $q$-series starting from modular differential equations up to some large finite power of $q$, and computation of the modular $\mS$-matrix using the contour integral representation. The first of these is clearly limited to finite powers of $q$ and for this reason our results (about both the existence of quasi-characters and the bilinear relation between them) are conjectural from a rigorous mathematical point of view. Proving the existence of quasi-characters in order 3 would require mathematical results analogous to those obtained in \cite{Kaneko:1998, Kaneko:2003, Kaneko:2006, Kaneko:2013uga} for the order 2 case.
A related point is that while the classification of admissible characters in order 2 is complete \cite{Chandra:2018pjq}, we make no claim to completeness here and this issue needs to be investigated further. 

We note that in \cite{Gaberdiel:2016zke} the bilinear relation among characters was proved using the hypergeometric representation of the characters and considerations of holomorphy. This was an exact result, not relying on the $q$-series expansion. An analogous result should be provable in order 3 using the contour-integral representation of (quasi-)characters as in \cite{Mukhi:1989qk,Mukhi:2019cpu}. This is definitely worthy of investigation. 

A complete classification of quasi-characters in order 3, together with an algorithm to combine them suitably, would complete the classification of admissible characters in order 3. 

\section*{Acknowledgements}

RP and PS would like to acknowledge the INSPIRE Scholarship for Higher Education, Government of India. All of us are grateful for support from a grant by Precision Wires India Ltd.\ for String Theory and Quantum Gravity research at IISER Pune. Part of this work was carried out in the congenial atmosphere of the National Strings Meeting at IISER Bhopal, December 2019, and we would like to thank the organisers for the kind hospitality and awesome food.

\section*{Appendix}

\appendix

\section{Verifying integrality and the bilinear relation}

\label{verifying}

Here we briefly summarise the procedure to find a recursive solution of the 3rd-order MLDE discussed in the body of this paper. More details may be found in \cite{Mathur:1988gt}. 
Thereafter we make a list of all the cases, as well as the order of expansion in powers of $q$, to which we have explicitly verified the results presented. 

The  conjectured quasi-characters must satisfy the following third-order MLDE with $\ell=0$:
\begin{equation}
  \left( {\mc D}^3 + \pi^2 \mu_1 E_4  {\mc D} + i\pi^3 \mu_2 E_6\right)\chi(\tau) = 0 \label{MLDE3}
\end{equation}
where $\mu_1, \mu_2$ are parameters, and $\mc D$ is $2\pi i $ times the covariant derivative in Eq. \eqref{MLDEgeneral}. 
By substituting the series expansion $\chi_i(\tau) = q^{\alpha_i} \sum_n a_n^{(i)} q^{n}$ in Eq. \eqref{MLDE3} and imposing the obtained indicial equation, one finds that the parameters are given by:
\begin{equation}
  \mu_1 = \frac{2}{9} - 4(\alpha_0 \alpha_1 + \alpha_0 \alpha_2 + \alpha_1 \alpha_2 );\quad
  \mu_2 = 8\alpha_0 \alpha_1 \alpha_2
\end{equation}
Thus the recurrence relation satisfied by the coefficients $a_k$ is
\begin{small}
\begin{equation}
  a_n^{(i)} = \frac{\sum_{s = 1}^n \Big(
    4(\alpha_i - s + n) \big( (2s-\alpha_i - n) E_{2,s} + 8 (\alpha_i \alpha_j + \alpha_i \alpha_k +\alpha_j \alpha_k )E_{4,s}\big) - 8 \alpha_i\alpha_j\alpha_k E_{6,s}\Big)a_{n-s}^{(i)}}
  {-2 n (1 + 2n - 2 \alpha_j - 4\alpha_k)(1 + 2n - 4\alpha_j - 2\alpha_k)} 
\end{equation}
\end{small}%
where $\alpha_i$ is the exponent of the character which is being solved for and  $\alpha_j, \alpha_k$ are the exponents of the other two characters.
In this expression $\alpha_j$ and $\alpha_k$ appear symmetrically and can be arbitrarily chosen to be either exponent. 

Clearly the solution is a function of the rational parameters $\alpha_i$, as described in the introduction.
To check which values of the exponents correspond to valid solutions as admissible characters or quasi-characters, we must check for integrality of the coefficients $a_n$, after suitably normalising the solution. 
This requires the denominators of the coefficients to stabilise beyond some finite order in $q$.
Hence, the solutions where the denominators do not appear to stabilise for large orders of $q$ are discarded.

It is important to note that the MLDE is very sensitive to the input parameters, i.e $c, h_i$. 
If one inputs values of $c, h_i$ which do not correspond to either admissible characters or quasi-characters, then the denominators grow rapidly with increasing orders in the $q$ expansion.
In fact, the LCM of the denominators, as a function of the order of expansion, is seen to grow exponentially. This is not the case for both characters and quasi-characters, where the denominator eventually stabilises to a fixed value. 

The existence of quasi-characters (namely the stabilisation referred to above) has been tested by us for the cases in Table \ref{verif-quasi}.

\begin{table}[h!]
  \centering
  \begin{tabular}{cccc}
    \toprule
    Family & $k$& $r$ & Order in $q$ \\
    \midrule
    $\mc M_{2,7}$ & $-50$ to 50 & - & 2000 \\
    \midrule
    \multirow{2}{*}{$\mc M_{2,5}\otimes \mc M_{2,5}$}  & $-50$ to 45 & - & 1000 \\
    &  45 to 50 & - & 1500 \\
    \midrule
    $G_{2,1} \otimes G_{2,1}$ & $-50$ to 50 & - & 1000 \\
    \midrule
    $F_{4,1} \otimes F_{4,1}$ & $-50$ to 50 & - & 1000 \\
    \midrule
    $A_{4,1}$ & $-50$ to 50; $~k\in\frac12\mathbb{Z}$ & - & 1000 \\
    \midrule
    $A_{2,1} \otimes A_{2,1}$ & $-50$ to 50; $~k\in\frac12\mathbb{Z}$ & - & 1000 \\
    \midrule
    $D_{r,1}$ & $-50$ to 50 & -20 to 20 & 500 \\
    \midrule
    $B_{r,1}$ & $-50$ to 50 & -20 to 20 & 500 \\
    \bottomrule
  \end{tabular}
  \caption{Range of verification of Quasi-character families}
  \label{verif-quasi}
\end{table}

The $q$-expansion was also used to check the bilinear relation between quasi-characters labelled by $k$ and $1-k$  in the same family.
A small caveat is that one needs to divide the bilinear product by $D_0 D'_0$ to obtain the Klein $j$-function plus a constant.
We have checked the bilinear relations to order $q^{10}$ and matched them with the $q$-expansion of $j$ for $k$ ranging from 1 to 10 and, where applicable, $r$ ranging from $-5$ to 5.

\bibliographystyle{JHEP}

\bibliography{quasi}

\end{document}